\pdfoutput=1

\documentclass[11pt]{article}
\usepackage{amssymb}
\usepackage[final]{acl}
\pdfoutput=1

\usepackage{acl}

\usepackage{times}
\usepackage{latexsym}
\usepackage{xspace}
\usepackage{color,xcolor}
\usepackage[T1]{fontenc}

\usepackage[utf8]{inputenc}
\usepackage{amssymb}
\usepackage{utfsym}
\usepackage{float}
\usepackage{pifont}

\usepackage{url}
\usepackage{enumitem}
\usepackage{booktabs}
\usepackage{multirow}
\usepackage{tabularx}
\usepackage{courier}
\usepackage{makecell}
\usepackage{amsmath}
\usepackage[most]{tcolorbox}
\usepackage{algorithm}
\usepackage{algpseudocode}
\usepackage{listings}
\usepackage{xcolor}
\usepackage{graphicx}
\usepackage{fancyvrb}
\usepackage{color}
\usepackage{subcaption} 
\usepackage{wrapfig}
\usepackage{multicol}
\definecolor{color_1}{HTML}{B8D9EB}
\definecolor{color_2}{HTML}{FEFF92}
\definecolor{color_3}{HTML}{99F0C2}

\usepackage{microtype}
\usepackage{inconsolata}

\newtcolorbox{tbox}[2][]{colback=yellow!6!white,colframe=blue!35!orange,left=3pt,right=3pt,top=-2pt,bottom=3pt,title=\small{#2},#1}
\usepackage{times}
\usepackage{latexsym}
\usepackage{array}
\usepackage[T1]{fontenc}

\usepackage{microtype}

\usepackage{inconsolata}

\usepackage{graphicx}
\usepackage{tabularx}
\title{EncGPT: A Multi-Agent Workflow for Dynamic Encryption Algorithms}

\author{
    \textbf{Donghe Li\textsuperscript{1,2}},
    \textbf{Zuchen Li\textsuperscript{3}},
    \textbf{Ye Yang\textsuperscript{1}},
    \textbf{Li Sun\textsuperscript{1}},
    \textbf{Dou An\textsuperscript{1}},
    \textbf{Qingyu Yang\textsuperscript{1,4}} \\
    \textsuperscript{1}Faculty of Electronic and Information Engineering, Xi'an Jiaotong University, Xi'an 710049, China \\
    \textsuperscript{2}MOE Key Laboratory for Intelligent Networks and Network Security, Xi'an Jiaotong University, Xi'an 710049, China \\
    \textsuperscript{3}School of Automation Science and Engineering, Xi’an Jiaotong University, Xi'an 710049, China \\
    \textsuperscript{4}SKLMSE Laboratory, Xi'an Jiaotong University, Xi'an 710049, China \\
    \small{\textbf{Corresponding author:} Qingyu Yang, \href{mailto:yangqingyu@mail.xjtu.edu.cn}{yangqingyu@mail.xjtu.edu.cn}} \\
}

\begin{document}
\maketitle
\begin{abstract}
Communication encryption has always been a critical infrastructure in computer technology. Existing encryption algorithms often struggle to balance cost and security—high security typically comes with high costs and low efficiency. Additionally, single encryption algorithms always carry security risks, and the design of dynamic algorithms lacks a feasible workflow. To address these issues, we propose EncGPT, a multi-agent workflow based on large language models (LLM). This framework consists of three main agents: the rule agent, encryption agent, and decryption agent. The rule agent dynamically generates encryption rules and features, while the encryption and decryption agents apply these rules and features to encrypt and decrypt the input, respectively. This successfully implements a dynamic encryption algorithm workflow for communication encryption, addressing the gap in LLM-MA for communication security. We tested GPT-4o’s encryption rule generation preferences and successfully implemented encryption-decryption workflows with homomorphism preservation in the substitution encryption algorithm. Our workflow achieved an average execution time of 15.99 seconds across the three substitution algorithms.
\end{abstract}

\section{Introduction}

With the ongoing development of information technology, encryption technology has become an essential cornerstone in the field of communication. Communication encryption ensures data privacy and protects against tampering through encryption and decryption processes. Initially used for military purposes, encryption has since expanded into a wide range of applications.\citep{lu1989secure,fieschi2023anonymization}

The origin of communication encryption dates back to ancient cryptography, such as the Caesar cipher and the Vigenère cipher\citep{luciano1987cryptology}. As computer technology developed, communication encryption gradually evolved into modern cryptographic techniques. It can broadly be divided into two stages\citep{diffie1988first,agrawal2012comparative}: symmetric encryption\citep{1089771,abdullah2017advanced}, which uses the same key for both encryption and decryption, and asymmetric encryption\citep{zhou2011research}, which employs two different keys. Although symmetric encryption offers good performance in terms of speed, it is no longer considered reliable today. If an attacker intercepts any of the keys, the security of the entire transmission is compromised\citep{zhang2021overview}. On the other hand, asymmetric encryption relies on complex mathematical challenges to ensure that attackers cannot deduce the private key from the public key. However, this comes with significant time and space complexity\citep{simmons1979symmetric}.Recent work has explored combining the strengths of both approaches by using asymmetric encryption to protect symmetric encryption keys\citep{dijesh2020enhancement}. However, no matter how these systems are improved, single encryption algorithms still present hidden security vulnerabilities. While some researchers have proposed dynamic encryption schemes, these are typically limited to the dynamic variation of keys or the chaotic generation of ciphertext\citep{ishaq2023dynamic,chai2017new}, rather than truly dynamic algorithm generation.

In conclusion, existing cryptographic algorithms have two major drawbacks:
1.They struggle to balance time, space costs, and encryption security. High security demands typically come with prohibitively high costs.
2.Single algorithms inherently carry security risks, and existing dynamic encryption technologies only enhance the dynamism of keys and ciphertext, not the algorithms themselves.Dynamic algorithm generation could be a feasible solution to these problems. To achieve this, a powerful agent system is required—not only to generate encryption rules and keys but also to encrypt and decrypt plaintext based on randomly generated encryption rules and keys.  Dynamic algorithm generation allows for the use of simple, inexpensive encryption rules for rapid multi-round encryption, enhancing security by keeping both the algorithm type and keys confidential.

\begin{figure*}[t]
\centering
  \includegraphics[width=0.9\linewidth]{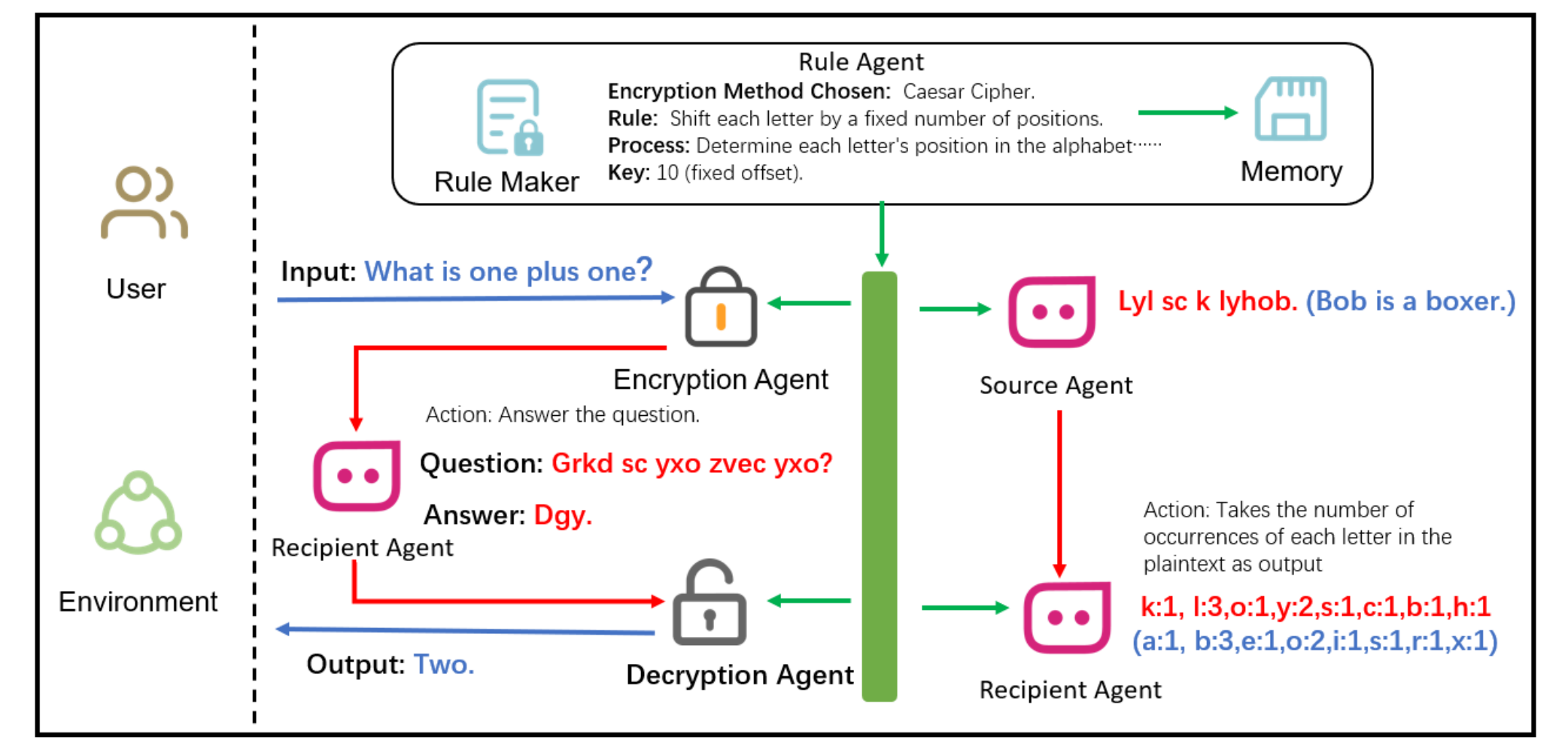} \hfill
  \caption {The architecture diagram of our workflow is shown. Green lines and the matrix represent the encrypted communication flow. On the left, the Recipient Agent interacts with the external environment, while on the right, agents communicate with each other. Red lines indicate the transmission of ciphertext, and blue lines indicate the transmission of plaintext.}
  \label{fig:workflow}
\end{figure*}

In recent studies, large language models (LLMs) based on deep neural network architectures have demonstrated exceptional potential in reasoning, planning, and creativity \citep{guo2024large}, particularly in multi-agent systems\citep{hong2023metagpt,zhou2023agents,DBLP:conf/acl/ZhangYH0XC24,DBLP:conf/acl/IslamAP24}. These systems assign different roles to various models and leverage collective intelligence to tackle complex tasks. Effective communication is crucial to maximizing the collective intelligence and unifying the objectives of a multi-agent system\citep{das2019tarmac}.In complex real-world tasks, multiple agents often need to collaborate, communicate, and reason over several rounds\citep{li2023camel}, making dynamic algorithm generation a highly suitable approach for such scenarios. However, we have discovered that research on the communication security of LLM-based multi-agent systems (LLM-MA) remains limited, which is a significant risk in the current development of LLM-MA. Therefore, in this paper, we propose using LLM-based multi-agent systems as a framework to implement dynamic encryption technology. We explore the application of communication encryption techniques within LLM-MA communication.

Building on this idea, we propose EncGPT, an encryption and decryption workflow based on the LLM-MA framework, designed for dialogue interaction scenarios within LLM-MA. Our workflow includes a rule agent, an encryption agent, and a decryption agent. The rule agent automatically generates encryption rules and algorithms in each round of dialogue and sends them into the encrypted communication flow. The encryption and decryption agents then encrypt and decrypt the input and output based on the generated rules and algorithms. The recipient agent responds to the ciphertext.

Our contributions are as follows:

\begin{itemize}
    \item We propose an agent-based communication encryption and decryption workflow, EncGPT. This workflow includes a rule agent, encryption agent, recipient agent, and decryption agent, allowing for the dynamic generation of encryption algorithms and keys, thereby enhancing encryption reliability.
    \item We discuss the communication security challenges in LLM-MA systems and present EncGPT as a potential solution to address these concerns.
    \item We designed an experiment to investigate the model’s preferences by generating encryption algorithm types using a Large Language Model (LLM). Additionally, we tested the encryption and decryption performance of GPT-4o and evaluated its homomorphic communication capabilities within the encryption workflow.
\end{itemize}

\section{Workflow Architecture}

Under the assumption of decentralized communication, we require that no plaintext appears in the agent communication flow except during the user input and output stages. The workflow architecture of EncGPT is illustrated in the figure\ref{fig:workflow}. During each communication round, the rule agent automatically generates an encryption algorithm and fills the key with a randomly generated number that complies with the algorithm. The rule agent then injects the encryption algorithm, including the key and other encryption parameters, into the encrypted communication flow, which generally maintains a higher level of security compared to the agent communication flow.

When a user or external environment communicates with the recipient agent, the encryption agent retrieves encryption parameters from the encrypted communication flow and encrypts the input. The ciphertext is then sent to the agent communication flow. The recipient agent processes the ciphertext and sends the encrypted response back. If the ciphertext is meant for the user or external environment, the decryption agent decrypts it using the encryption parameters and converts it into plaintext before outputting. For agent-to-agent communication, the recipient agent utilizes the ciphertext from the agent communication flow along with the encryption parameters from the encrypted communication flow to perform its designated actions. Additionally, all historical records of agents within the workflow must be cleared after each communication round. This measure prevents "prompt contamination" \citep{bender2021dangers},which could result from previous encryption rules interfering with subsequent interactions.

\section{Encryption Rules}

 we used classical encryption algorithms as the basis for generating rules through the agent. Classical encryption algorithms are less complex, computationally simpler, and have smaller keys, making them far more cost-effective than the other two types. Although most classical encryption algorithms can be broken using frequency analysis based on text\citep{carter2007classical}, they can still be effective in multi-round LLM-MA scenarios due to the confidentiality of the encryption algorithm type and cost considerations.

We use LLMs as the core of the agent, employing multi-round dialogue prompts to establish the encryption rules. In the first round, we ask the LLM to generate an encryption algorithm and use a mask to replace any numbers in the rule, including the key. In the following step, we require the LLM to define the common range for the masked numbers. In the final round, the model generates random numbers within this range and incorporates them into the encryption rules. This approach mitigates the LLM's "contextual bias," which causes it to repeatedly rely on specific numbers when directly generating rules. Additionally, the rule agent includes a memory module to record the encryption rules and keys.

\section{Encryption and Decryption}
Both the encryption and decryption agents are built around the LLM core. For their construction, we utilize chain-of-thought\citep{wei2022chain} reasoning as prompts. 

\section{Homomorphism}

The concept of homomorphism in encryption was first introduced in 1978 \citep{rivest1978data}. The core idea is to perform operations on encrypted data without exposing the sensitive information. We aim to use the LLM to directly execute this process. In our experiment, we explore the homomorphic properties of the LLM.

\section{Experiment}

All prompt templates used in our study are provided in the appendix\ref{sec:prompt}.
\subsection{Rule Preferences}
\begin{figure}[t]
  \includegraphics[width=\columnwidth]{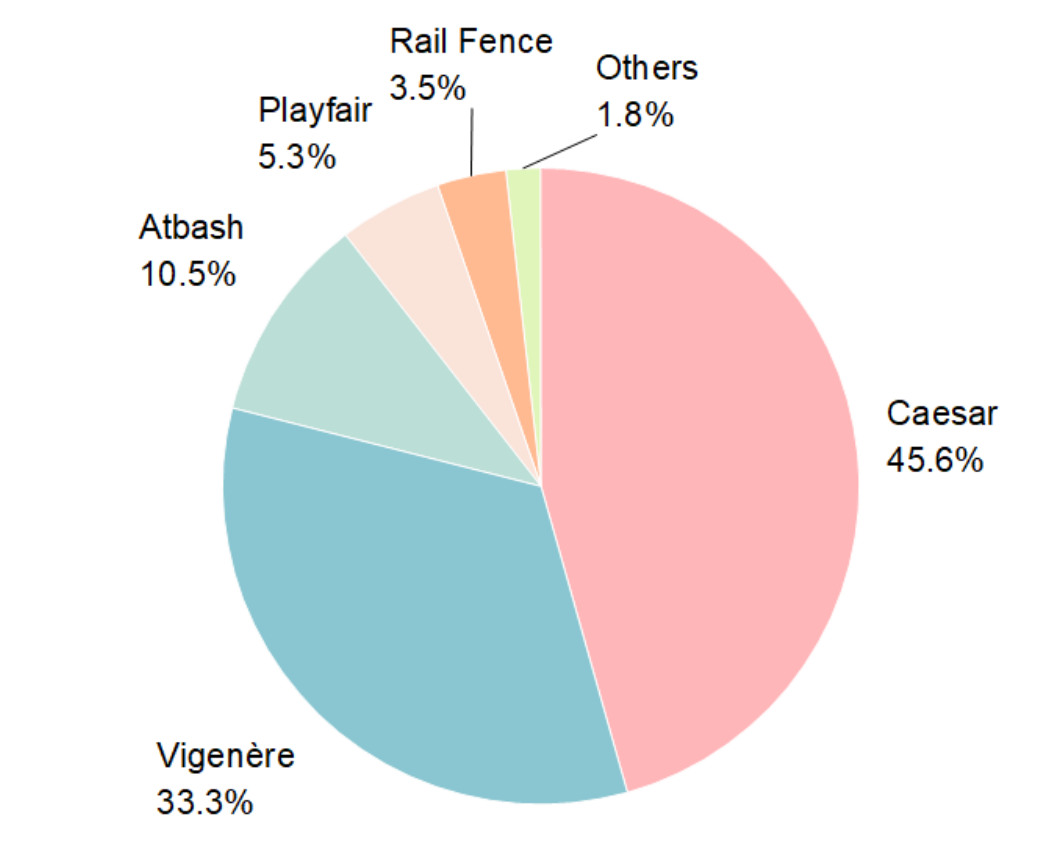}
  \caption{GPT-4o’s Preferences in Encryption Rule Generation}
  \label{fig:favor}
\end{figure}

We used GPT-4o as the rule agent's generator and analyzed its algorithm selection preferences. The distribution of selected algorithms is illustrated in Figure\ref{fig:favor}.

\begin{table*}[ht]
  \centering
  \begin{tabular}{ccccccc}
    \hline
    \          & \textbf{Caesar} & \textbf{Vigenère} & \textbf{Atbash} & \textbf{Playfair} & \textbf{RailFence} & \textbf{Others} \\
    \hline
    E-D & \checkmark & \checkmark & \checkmark & \texttimes & \texttimes & \texttimes \\

    E-R-D & \checkmark & \checkmark & \checkmark & \texttimes & \texttimes & \texttimes \\
    \hline
  \end{tabular}
  \caption{\label{table1}
    Encryption and decryption of various algorithms
  }
\end{table*}

In the model preference survey, GPT-4o clearly favored the Caesar and Vigenère algorithms, which was not our intended outcome, despite prompting the model to exercise creativity in the instructions.
\subsection{Encryption and Decryption}

\begin{table}[ht]
  \centering
  \begin{tabular}{ccccc}
    \hline
    \            & \textbf{Rule Gen} & \textbf{Enc } & \textbf{Dec} &\textbf{Total} \\
    \hline
    \textbf{Caesar}                  & 13.36s                    & 6.45s                & 7.51s                 & 27.23s                             \\ 
\textbf{Vigenère}                & 14.13s                   & 8.42s                & 8.34s                 & 30.89s                             \\ 
\textbf{Atbash}                  & 10.44s                    & 6.57s                & 7.32s                 & 24.33s                             \\ 
\hline
  \end{tabular}
  \caption{\label{table2}
    Time Consumption of Each Part of EncGPT and the Total Workflow Time
  }
\end{table}

We tested GPT-4o's capabilities as both an encryption agent and a decryption agent. For simplicity, we temporarily removed the receiver agent and treated the "encryption-decryption" process as a single method. If the ciphertext encrypted by the encryption agent is successfully decrypted by the decryption agent to return the correct plaintext, we consider the encryption-decryption process successful. We collected data on GPT-4o's performance with various encryption algorithms.

Additionally, we treated the "encryption-receiver agent-decryption" process as a method, where the task for the receiver agent was defined as: "Please organize the encrypted text, decrypt it according to the encryption rules, and output the frequency of each letter in the plaintext." The success or failure of the receiver agent in this task serves as a critical test of whether the LLM can maintain homomorphism within the encryption-decryption workflow.

As shown in Table\ref{table1},GPT-4o performed well with character substitution algorithms such as Caesar,Vigenère, and Atbash, demonstrating both effective decryption and the ability to maintain homomorphism. We carefully analyzed the experimental data for other algorithms and attempted to summarize the reasons for these results. While generating rules, GPT-4o inevitably produced "hallucinations\citep{ji2023survey}", where the encryption rules for some data were incorrect. Additionally, GPT-4o's mathematical capabilities are limited, making it difficult or impossible to solve complex tasks such as matrix recognition, and other advanced mathematical logic problems.

As shown in Table\ref{table2}, we measured the execution time of each component and the total time for the EncGPT workflow under the replacement algorithm. The average execution time across the three algorithms was 15.99 seconds.

\section{Conclusion}
This paper presents EncGPT, a communication encryption method based on the LLM-MA system, leveraging the creativity and reasoning capabilities of LLMs to implement rule generation and encryption-decryption workflows. This approach addresses the conflict between cost and security in existing algorithms, as well as the security risks associated with single encryption algorithms. The workflow consists of a rule agent, an encryption agent, and a decryption agent, each responsible for generating algorithms and executing encryption and decryption activities. Experimental results show that GPT-4o has a preference for the character substitution algorithm and successfully implemented the workflow using this algorithm. The average execution time for the three substitution algorithms was 15.99 seconds, indicating room for optimization compared to traditional heuristic algorithms.

Overall, our workflow represents the first attempt to use LLMs to generate dynamic encryption rules, filling a gap in the LLM-MA field’s research on communication security. Future work will focus on enhancing agent capabilities, including the use of tools, fine-tuning, and the implementation of Retrieval-Augmented Generation(RAG)\citep{lewis2020retrieval} techniques. Additionally, we will aim to further reduce time and space costs by quantifying the model's performance.

\section{Limitations}
A major limitation of our work is the insufficient exploration of LLM-Agent capabilities. As demonstrated in the experiments, we were unable to implement effective measures to address issues such as GPT-4o's algorithmic bias, 'hallucinations,' underperformance in solving mathematical or special logic problems. These issues affect the model’s rule generation randomness and execution accuracy. Future work should focus on enhancing LLM-Agent capabilities and further reducing time and space costs. Additionally, this paper does not conduct an in-depth investigation of symmetric encryption algorithms. However, our researchers performed simple tests, and current LLMs are unable to achieve satisfactory results with symmetric encryption. Integrating symmetric encryption algorithms into the workflow will be a key focus of future work.

Another limitation is the lack of exploration into communication signatures and receiver identification. While this area remains unexplored in the LLM-MA framework, there are existing solutions in prior multi-agent systems research. For example,previous researchers\citep{das2019tarmac} have proposed a signature-based soft attention mechanism for identifying receiver agents in communication. Future research may draw inspiration from their approach to address similar issues.


\appendix

\section{Prompt}
\label{sec:prompt}
\subsection{Rule Agent}
In Table\ref{table3}, we will present the prompts used for the rule agent. These prompts are designed for a three-round dialogue.

\begin{table}[ht]
    \centering
    \begin{tabular}{|p{\columnwidth}|}
        \hline
        \textit{You are an expert in creating encryption rules. Your task is to design a specific encryption scheme for natural language. First, select an encryption rule. The rule can be a classical encryption method. You are encouraged to use your imagination, but you can only select one encryption rule. In your response, only mention the chosen algorithm and the encryption process. All numbers involved in your explanation must be represented by a mask. Please use simple language. Please use the following format for your response:} \\
        \textbf{Encryption Method Chosen:}\textit{The type of encryption rule you choose }  \\
        \textbf{Rule:}\textit{Rules of the encryption algorithm}   \\
        \textbf{Process:}\textit{Specific steps of encryption}   \\
        \textbf{Key:}\textit{Key or other specific encryption characteristics and details}   \\
        \hline
        \textit{Great job! Now, think about the possible range and values of all the numbers represented by masks in your chosen encryption rule.}\\
        \hline
        \textit{Well done! Next, please randomly generate numbers within the specified range, fill the selected numbers into the encryption rules you’ve established, and generate the complete encryption rule. Output the result using the following format:} \\
        \textbf{Encryption Method Chosen:} \\
        \textbf{Rule:}   \\
        \textbf{Process:}  \\
        \textbf{Key:}  \\
        \hline
    \end{tabular}
    \caption{Prompt for rule agent}
    \label{table3}
\end{table}

\subsection{Encryption Agent}
In table\ref{table4}, we present the prompts used for the encryption agent. We need to use a heuristic filler to populate the content inside the \{\} brackets, which includes the encryption rules and algorithm features generated by the rule agent, as well as the input plaintext. The encryption rules are retrieved by the encryption agent from the encrypted communication flow.
\begin{table}[ht]
    \centering
    \begin{tabular}{|p{\columnwidth}|}
        \hline
        \textit{You are a natural language encryption expert. Your task is to encrypt my plain text based on the encryption rules. Carefully read and understand each encryption rule, and for each rule, you need to convert it into the corresponding encryption method. Then, use these methods to encrypt the plaintext and output the ciphertext.} \\
        \textbf{Encryption Rules:}\{\}  \\
        \textbf{Plaintext:}\{\}   \\
        \textit{Your answer should follow this format:}\\
        \textbf{Reasoning Process: }  \\
        \textbf{Ciphertext Answer:}  \\
        \hline
    \end{tabular}
    \caption{Prompt for encryption agent}
    \label{table4}
\end{table}

\subsection{Decryption Agent}
In table\ref{table5}, we present the prompts used for the decryption agent. We need to use a heuristic filler to populate the content inside the {} brackets, which includes the encryption rules and algorithm features generated by the rule agent, as well as the output ciphertext. The encryption rules are retrieved by the decryption agent from the encrypted communication flow.
\begin{table}[ht]
    \centering
    \begin{tabular}{|p{\columnwidth}|}
        \hline
        \textit{You are a decryption expert, and your task is to decrypt my ciphertext based on the provided encryption rules. Carefully review each rule, and for every rule, devise an appropriate decryption method. Once you have identified the decryption methods, apply them to the ciphertext to produce the plaintext.} \\
        \textbf{Encryption Rules:}\{\}  \\
        \textbf{Ciphertext:}\{\}   \\
        \textit{Your answer should follow this format:}\\
        \textbf{Reasoning Process: }  \\
        \textbf{Plaintext Answer}  \\
        \hline
    \end{tabular}
    \caption{Prompt for decryption agent}
    \label{table5}
\end{table}

\subsection{Agent Communication}
In Table\ref{table6}, we present the prompts used for communication between agents. We need to use a heuristic filler to populate the content inside the \{\} brackets, which includes the recipient agent's original task, the encryption rules and algorithm features generated by the rule agent, and the input ciphertext. The encryption rules are retrieved by the decryption agent from the encrypted communication flow.
\begin{table}
    \centering
    \begin{tabular}{|p{\columnwidth}|}
        \hline
        \textit{You are an encryption and decryption expert, your task is to restore the ciphertext input according to the encryption rules, then operate on the plaintext, and finally need to use the encryption rules to encrypt your results.} \\
        \textbf{Encryption Rules:}\{\}  \\
        \textbf{Ciphertext input:}\{\}   \\
        \textit{At the same time, you are also \{a letter statistician\}, your task is to \{count each letter in the plaintext and the number of times it appears\}}\\
        \textit{Your output should also be encrypted using encryption rules.
        Your task chain is: first decrypt my ciphertext input, next, perform \{letter statistics\}, and finally encrypt and output your results.
        Your output should be in the following format:}\\
        \textbf{Decryption Thinking:} \\
        \textbf{Enter plaintext:} \\
        \textbf{Working on plaintext:} \\
        \textbf{Work result:} \\
        \textbf{Crypto thinking: }\\
        \textbf{Encrypted output: }\\
        \hline
    \end{tabular}
    \caption{Prompt for agent homomorphic communication}
    \label{table6}
\end{table}

\section{Case Study}
\begin{figure*}[t]
\centering
  \includegraphics[width=0.9\linewidth]{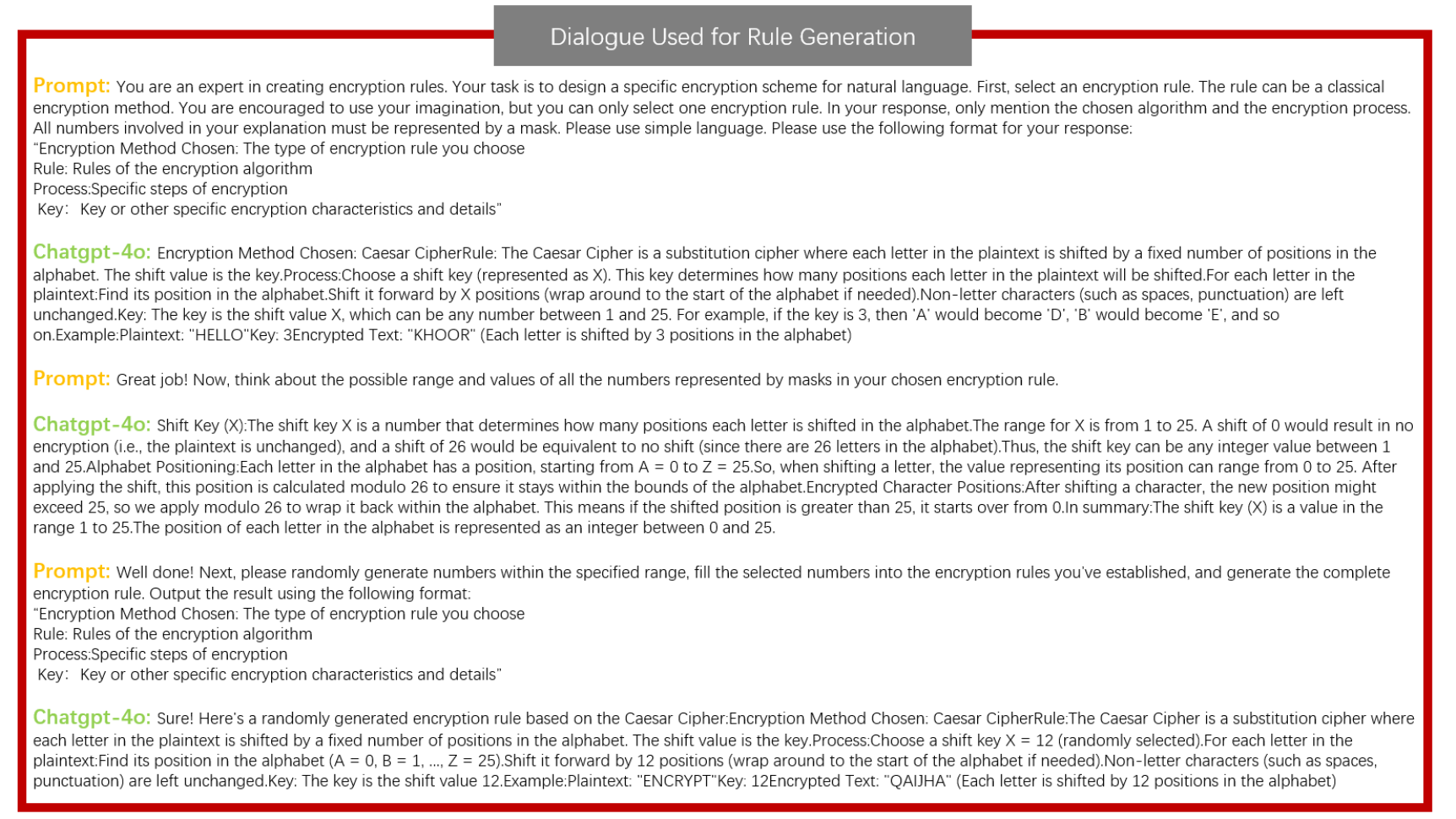} \hfill
  \caption {Dialogue Used for Rule Generation
}
  \label{fig:rule}
\end{figure*}

\begin{figure*}[t]
\centering
  \includegraphics[width=0.9\linewidth]{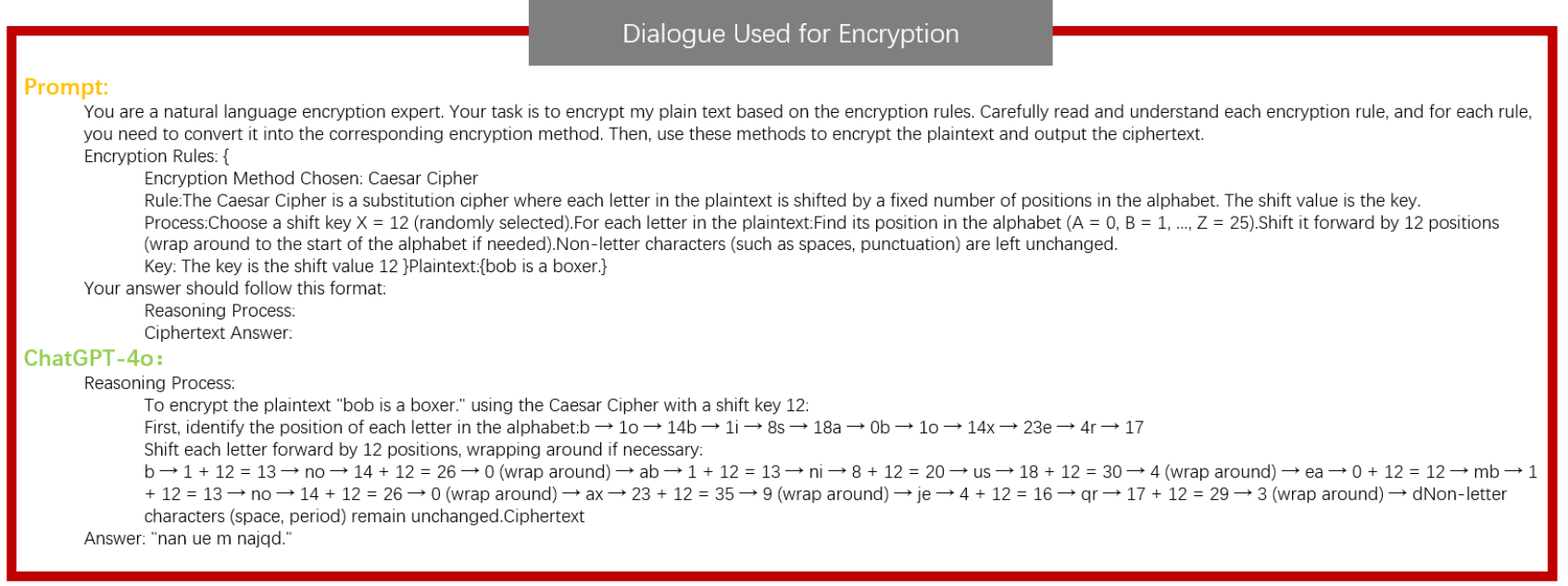} \hfill
  \caption {Dialogue Used for Encryption }
  \label{fig:enc}
\end{figure*}

\begin{figure*}[t]
\centering
  \includegraphics[width=0.9\linewidth]{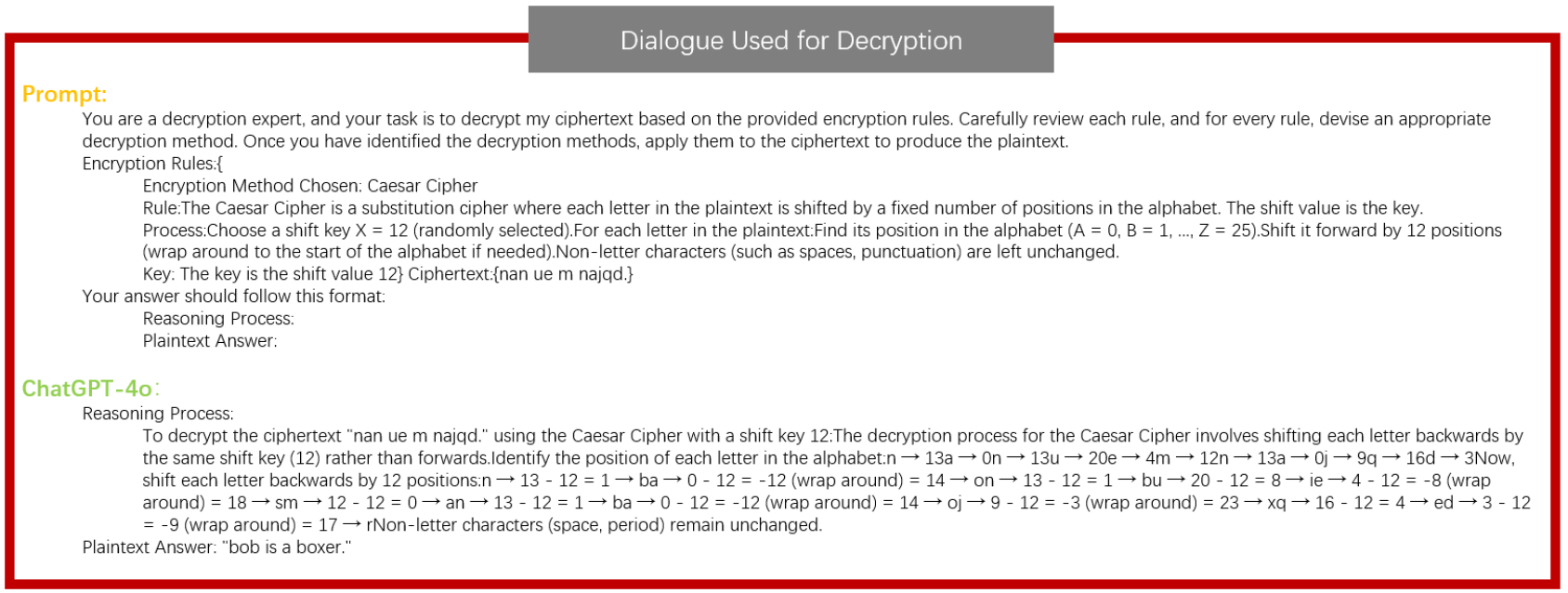} \hfill
  \caption {Dialogue Used for Decryption }
  \label{fig:dec}
\end{figure*}

\begin{figure*}[!htbp]
\centering
  \includegraphics[width=0.9\linewidth]{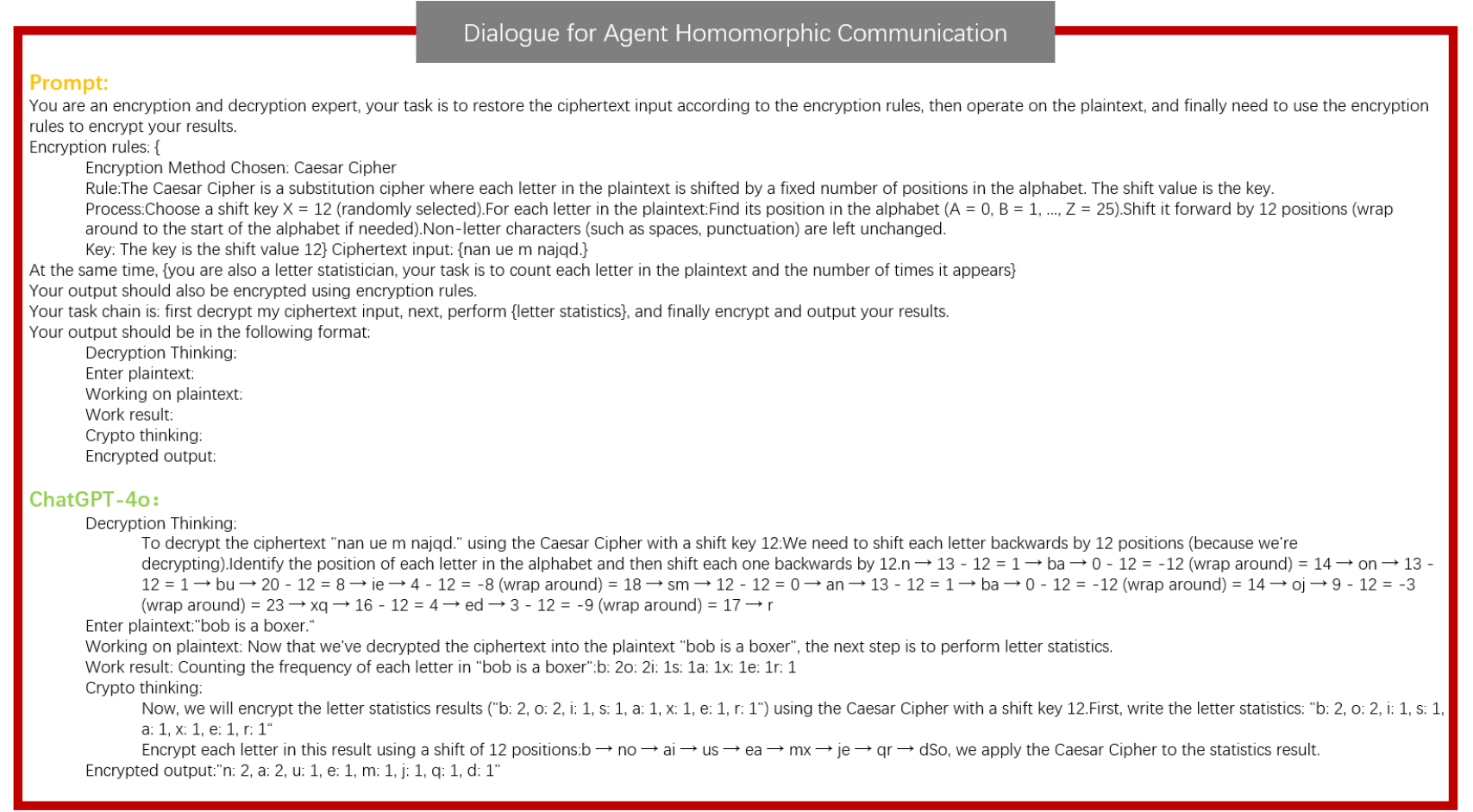} \hfill
  \caption {Dialogue for Agent Homomorphic Communication}
  \label{fig:agcom}
\end{figure*}

Figures \ref{fig:rule},\ref{fig:enc},\ref{fig:dec},\ref{fig:agcom} showcase the applicable cases during our testing. In this case, GPT-4o generated the Caesar cipher algorithm and its features, successfully implementing encryption, decryption, and homomorphic communication.

\end{document}